\documentclass{emulateapj}

\usepackage{apjfonts}
\usepackage{CJK}
\usepackage{natbib}
\def\msun {\rm M_{\sun}}
\def\lsun {\rm L_{\sun}}

\begin{document}
\begin{CJK}{GB}{gbsn}

\shorttitle{SED of Dusty Quasars}
\shortauthors{Dai et al.}

\title{A Population of Dust-Rich Quasars at {\it z} $\sim$ 1.5}
\author{Y. Sophia Dai({´÷êÅ})\altaffilmark{1,2}, 
Jacqueline Bergeron\altaffilmark{3},
Martin Elvis\altaffilmark{1},
Alain Omont\altaffilmark{3},
Jia-Sheng Huang\altaffilmark{1}, 
Jamie Bock\altaffilmark{4,5},
Asantha Cooray\altaffilmark{6},
Giovanni Fazio\altaffilmark{1},
Evanthia Hatziminaoglou\altaffilmark{7},
Edo Ibar\altaffilmark{8,9},
Georgios E. Magdis\altaffilmark{10,18},
Seb J. Oliver\altaffilmark{11},
Mathew J. Page\altaffilmark{12},
Ismael Perez-Fournon\altaffilmark{13,14},
Dimitra Rigopoulou\altaffilmark{10,15},
Isaac G. Roseboom\altaffilmark{9},
Douglas Scott\altaffilmark{16},
Myrto Symeonidis\altaffilmark{12},
Markos Trichas\altaffilmark{1},
Joaquin D. Vieira\altaffilmark{4},
Christopher N. A. Willmer\altaffilmark{17},
and Michael Zemcov\altaffilmark{4,5}
 }

\altaffiltext{1}{Harvard-Smithsonian Center for Astrophysics, 60 Garden Street,
Cambridge, MA 02138, USA; ydai@cfa.harvard.edu}
\altaffiltext{2}{Boston College, 140 Commonwealth Ave, Chestnut Hill, MA 02468, USA}
\altaffiltext{3}{Institut d'Astrophysique de Paris, UMR7095, 98bis boulevard Arago, F-75014, Paris, France}
\altaffiltext{4}{California Institution of Technology, 1200 East California Blvd, Pasadena CA 91125}
\altaffiltext{5}{Jet Propulsion Laboratory, 4800 Oak Grove Drive, Pasadena, CA 91109, USA}
\altaffiltext{6}{Dept. of Physics and Astronomy, University of California, Irvine, CA 92697-4575, USA} 
\altaffiltext{7}{ESO, Karl-Schwarzschild-Str. 2, 85748 Garching bei M\"unchen, Germany}
\altaffiltext{8}{UK Astronomy Technology Centre, Royal Observatory, Blackford Hill, Edinburgh EH9 3HJ, UK}
\altaffiltext{9}{Institute for Astronomy, University of Edinburgh, Royal Observatory, Blackford Hill, Edinburgh EH9 3HJ, UK}
\altaffiltext{10}{Dept. of Astrophysics, Oxford University, Keble Road, Oxford OX1 3RH, UK}
\altaffiltext{11}{Astronomy Centre, Department of Physics and Astronomy, University of Sussex, Brighton BN1 9QH}
\altaffiltext{12}{Mullard Space Science Laboratory, University College London, Holmbury St Mary, Dorking, Surrey RH5 6NT}
\altaffiltext{13}{Instituto de Astrof\'isica de Canarias (IAC), E-38200 La Laguna, Tenerife, Spain}
\altaffiltext{14}{Departamento de Astrof\'isica, Universidad de La Laguna (ULL), E-38205 La Laguna, Tenerife, Spain}
\altaffiltext{15}{Space Science \& Technology Dept., Rutherford Appleton Laboratory, Chilton, Didcot, Oxfordshire OX11 0QX, UK}
\altaffiltext{16}{Dept. of Physics and Astronomy, University of British Columbia, Vancouver, B.C., V6T 1Z1, Canada}
\altaffiltext{17}{Steward Observatory, University of Arizona, 933 North Cherry Avenue, Tucson, AZ 85721, USA}
\altaffiltext{18}{CNRS/CEA Saclay, Service d\'Astrophysique Orme des Merisiers, F-91191 Gif-sur-Yvette Cedex, France}

\begin{abstract}
We report {\it Herschel}\footnote{ {\it Herschel} is a ESA space observatory with science 
 instruments provided by European-led Principal Investigator consortia and with important 
 participation from NASA.} 
 SPIRE (250, 350, and 500\,$\mu$m) detections of 32 quasars with redshifts
0.5 $\leq z <$ 3.6 from the {\it Herschel} Multi-tiered Extragalactic Survey (HerMES\footnote{http://hermes.sussex.ac.uk}). 
These sources are from a MIPS 24\,$\mu$m flux-limited sample of 326 quasars in 
the Lockman Hole Field.
The extensive multi-wavelength data available in the field permit construction of the 
rest-frame Spectral Energy Distributions (SEDs)
from ultraviolet to the mid-infrared for all sources, 
and to the far-infrared (FIR) for the 32 objects. 
Most quasars with {\it Herschel} FIR detections 
show dust temperatures in the range of 25\,K to 60\,K, with a mean of 34K. 
The FIR luminosities range from $10^{11.3}$ to $10^{13.5}\,$L$_{\sun}$, 
qualifying most of their hosts as ultra- or hyper-luminous infrared galaxies.
These FIR-detected quasars may represent a dust-rich population, 
but with lower redshifts and fainter luminosities than quasars observed at $\sim$ 1\,mm.
However, their FIR properties cannot be predicted from shorter wavelengths (0.3--20\,$\mu$m, rest-frame),
and the bolometric luminosities derived using the
5100\,\AA\ index may be underestimated for these FIR-detected quasars.
Regardless of redshift, we observed a decline in the relative strength of FIR luminosities 
for quasars with higher near-infrared luminosities.
\end{abstract}


\section{INTRODUCTION}
\label{sec:intro}
The far-infrared (FIR) properties of quasars are critical for our understanding of active galaxy evolution,
as they probe the possible evolutionary connection between star formation
and black hole (BH) accretion \citep[e.g.][]{sanders88, lutz07}.
Both starbursts and active galactic nuclei (AGN) could contribute to the rest-frame FIR emissions
at various redshifts, 
although starbursts are considered to be dominant \citep[e.g.][]{rr95, trichas09}.
Other models have successfully explained FIR luminosity as originating from direct AGN heating,
where cool dust at large distances from the AGN reside in a warped disk/torus \citep[e.g.][]{sanders89,haas03}.
Spectral information can be used to break this degeneracy. 
Polycyclic Aromatic Hydrocarbon (PAH) features, for instance, are used to indicate star formation activity \citep[e.g.][]{lutz07}.
{\it Spitzer} spectroscopic studies detected PAH emission in some of
the Palomar-Green (PG) and Sloan Digital Sky Survey (SDSS) quasars, where strong star formation coexists with quasars, 
and is responsible for an average of 60\% of the FIR emission \citep[e.g.][]{hao05, netzer07, lutz08, veilleux09, shi09}.

Rest-frame FIR SED studies provide powerful constraints on the star formation in quasars. 
Despite large dispersions at different wavelengths for optically bright, unobscured quasars, 
their mean SEDs show surprising uniformity over redshift, luminosity, and Eddington ratio
\citep[e.g.][]{elvis94, rich06, hao11}.
However, the SEDs at rest-frame $\lambda$ $>$ 40\,$\mu$m
were poorly defined at high redshifts 
(IRAS 100\,$\mu$m was the longest wavelength for E94 and R06).
Several groups have tried to address this FIR gap 
with various sample selections \citep[e.g.][]{papo06, kartaltepe10}; 
others have taken advantage of (sub)millimeter observations, 
e.g., with the Infrared Space Observatory (ISO), the Institut de Radioastronomie Millim$\acute{e}$trique (IRAM)
telescope, and the Submillimetre Common-User Bolometer Array (SCUBA) \citep[e.g.][]{omont01, omont03, haas03, priddey03}.
However, a single photometric point in the rest-frame FIR, 
as is the case for most (sub)mm quasar studies,
does not strongly constrain the dust temperature distribution
for these quasars. 
The sample sizes for (sub)mm quasars with enough photometric points
that allow FIR SED studies, on the other hand,
are limited due to the relatively long exposure times required for detections
\citep[e.g.][]{bee06, wang08, wang10}. Only about 10 (sub)mm quasars
reported to date have detailed rest-frame FIR SED measurements.

The {\it Herschel} Space Observatory \citep{pilbratt10} 
has opened a new window \citep{grif10} to directly study the rest-frame FIR properties
for quasars with moderate redshifts ($z \sim$ 1.5).  
The {\it Herschel} Multi-tiered Extragalactic Survey \citep{oliver10, oliver11}
covers $\sim$ 70 deg$^2$ with rich multi-wavelength data. 
Along with other {\it Herschel} surveys,
rest-frame 30--300\,$\mu$m emissions, likely due to cold dust 
in quasars and other AGNs, have been detected \citep{hatzi10, leip10, serjeant10}.

In this paper, we report 32 {\it Herschel} SPIRE detections of 24\,$\mu$m flux-limited 
quasars in the HerMES Lockman Hole field.  
This SPIRE--detected sample constitute at least 10\% of the 24\,$\mu$m selected quasar sample in this field,
and allow construction of the complete FIR SEDs for 
these broad-line quasars at $ z  \sim$ 1.5, 
which triples the size of (sub)mm observed quasars that have detailed FIR SEDs.
Throughout the paper, we assume a concordance cosmology with H$_0$=70 km$/$s Mpc$^{-1}$, $\Omega_{\rm M}$=0.3, and
$\Omega_{\Lambda}$=0.7.  

\section{MIPS 24\,$\mu$m-SELECTED QUASARS AND THEIR FIR COUNTERPARTS}
\label{sec:observation}
The quasars used in this paper are from a 24\,$\mu$m flux-limited
sample in the {\it Spitzer} Wide-area InfraRed Extragalactic Survey \citep{lonsdale03}.  
In the Lockman Hole--SWIRE (LHS) field,
we selected targets that satisfy MIPS $S_{24} >$ 0.4\,mJy ($\sim$ 8\,$\sigma$), 
and 94\% of the flux-limited sample also satisfy SDSS $r_{\rm AB}$ $<$ 22.5. 
In 2009, Huang et al (in prep.) performed a spectroscopic survey of $\sim$ 3000 such 24\,$\mu$m targets
with HECTOSPEC \citep{fab05} on the Multiple Mirror Telescope (MMT), with an effective coverage of $\sim$ 8\,deg$^2$. 
93\% of these objects have reliable redshifts.
SDSS objects with existing spectroscopic $z$ \citep{hatzi08} that satisfy the same flux-limits
were later added to the 24\,$\mu$m flux-limited sample, 
which increased the spectroscopic completeness to $\sim$ 70\%.
Broad line quasars were then selected, where MgII or CIV line width has a FWHM $>$ 1000\,km s$^{-1}$ \citep{sch07}.
The final sample of 326 24\,$\mu$m-selected sources includes 210 MMT and 116 SDSS quasars.

\begin{figure}[h]
\begin{center}
\includegraphics[scale=0.7,angle=0]{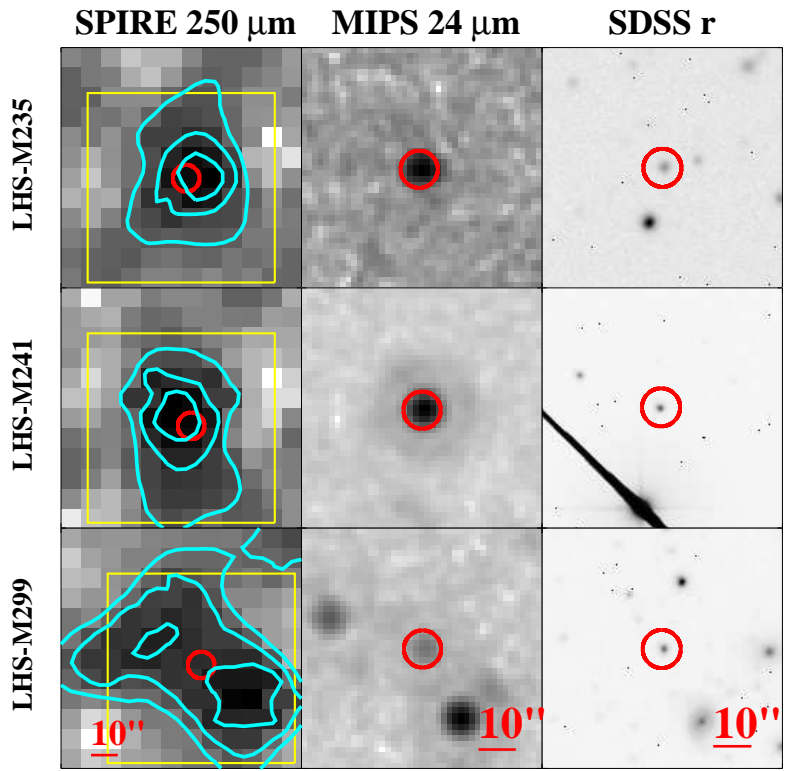}
\end{center}
\figcaption{Multi-wavelength images for two FIR-detected (top and middle) and one confused (bottom) quasars.
The red circles in all three bands indicate the optical positions and have a 5\,$\arcsec$ radius. 
The sizes of the optical-mid-IR images are 50\,$\arcsec$ $\times$ 50\,$\arcsec$,  corresponding to the yellow squares in the 250\,$\mu$m images.  
The cyan contours in the 250\,$\mu$m image mark the 3, 6, and 9 $\times$ $\sigma$.
The FIR emission corresponds to an isolated 24\,$\mu$m source in all the 32 FIR-detected quasars listed in Table.~\ref{tab:param}. 
The bottom panel shows one of the four confusing cases excluded from the 41 sources with 5$\sigma$ SPIRE signal. 
\label{fig:stamp}}
\end{figure}

We matched these 326 quasars to the {\it HerMES} SPIRE cross-identification (XID) catalog \citep{roseboom10}.
The XID catalog used SWIRE MIPS 24\,$\mu$m positions to minimize the source blending effects
due to large beam sizes (18$\arcsec$ 
FWHM for SPIRE 250\,$\mu$m images),
and has a completeness of $\sim$ 80\% at $S_{250} =$ 20mJy. 
Among the 326 24\,$\mu$m-selected quasars, there are 41 SPIRE detections with S/N $>$ 5, 
of which three were detected at  350 or 500\,$\mu$m only. 
We dropped four sources whose SPIRE 250\,$\mu$m beam
covers two 24\,$\mu$m counterparts (e.g. Fig.~\ref{fig:stamp}, bottom).
We also excluded the five {\it z} $<$ 0.5 objects 
because their rest-frame FIR data points 
do not constrain the SED fitting.
The final sample consists of 32 quasars (20 MMT and 12 SDSS objects) at 0.50 $\leq$ {\it z} $\leq$ 3.54,
with a median {\it z} of 1.55 (Fig.~\ref{fig:submm}, inset).
This corresponds to a 10$\%$ detection rate.
Since 29 of them were SPIRE 250\,$\mu$m detected,
the 32 quasars used in this paper are hereafter referred to as FIR-detected quasars (Fig.~\ref{fig:stamp}, top and middle panels).
Twenty-seven sources have at least one 350 or 500\,$\mu$m detection ($>$ 3$\sigma$), 
and 16 sources also have SWIRE MIPS 70 or 160\,$\mu$m detections.

\section{SPECTRAL ENERGY DISTRIBUTION}
\label{sec:SED}
We constructed the rest-frame SEDs for the 32 FIR-detected quasars from the UV to the FIR bands. 
The LHS field was covered by 
the Galaxy Evolution Explorer (GALEX) in the ultra-violet ({\em FUV, NUV}),
SDSS in the optical (u, g, r, i, and z),  
the UKIRT Infrared Deep Sky Survey (UKIDSS) 
in the near infrared (NIR; {\em J,H,K}), and the SWIRE survey in the mid-infrared 
(IRAC at 3.6, 4.5, 5.8, and 8.0\,$\mu$m; MIPS at 24, 70, and 160\,$\mu$m). 
A Chandra X-ray survey covered a small fraction of the LHS field (0.7 deg$^2$),
and only one source (LHS-S119) was detected out of the three FIR-detected quasars within that area \citep{wilkes09}.
Fig.~\ref{fig:stamp} shows the stamp images for two FIR-detected quasars and 
one poorly matched quasar in the optical and infrared bands.

The rest-frame FIR emissions of these quasars fall in 
the 30--300\,$\mu$m region, similar to those of ISO observed PG quasars \citep{haas03}
and (sub)mm-detected quasars
\citep[e.g.][]{mcmahon99,willott03, priddey03, priddey03b, robson04, 
omont01, omont03, car01, bee06, wang07,wang08,wang10}.  
 Using the same method as was adopted in \citet{bee06}\footnote{$M_{\rm d} = 
S_{\rm \nu0} {D_{\rm L}}^2 / (1+z) k_{\rm d}(\nu) B(\nu,T_{\rm d})$,
where $k_{\rm d}(\nu) = k_0  (\nu / \nu_{0})^{\beta}$ is the dust absorption coefficient.
Here we used $S_{\rm 250}$, and $k_d$ is from \citet{alton04}.}, 
we derived the dust mass $M_{\rm d}$ for the FIR-detected quasars.
$M_{\rm d}$ is in the order of 10$^8$--10$^9$\,M$_{\sun}$, 
similar to that of (sub)mm quasars.
These values are 1$\sim$2 dex higher than values estimated for the PG quasars. 
Therefore, these FIR-detected quasars are associated with the (sub)mm detected quasars
as a `dust-rich quasar' population (Fig.~\ref{fig:submm}).

In 31 out of the 32 quasars, the rest-frame SEDs exhibit FIR excess over the E94/R06 quasar 
templates (Sec.~\ref{sec:intro}) by 0.5 to 2.3 dex at 90\,$\mu$m, with an average of 1.4\,dex. 
This suggests that the contribution from cool dust is present. 
We compared the mean SEDs for FIR-detected and the much larger sample of undetected quasars (Fig.~\ref{fig:sedb}). 
These mean SEDs were constructed by combining individual rest-frame SEDs. 
We first converted the flux densities to luminosities for each object.  
After shifting their bandpasses to the rest frame, we normalized each SED at 1--5\,$\mu$m to the R06 template. 
We then populated a grid of points separated by 0.03 in log wavelength and 
linearly interpolated between the effective detections from the UV to the FIR.
The mean luminosities with this gap-repaired photometry were then connected as the average SED (magenta curves).
For FIR-undetected objects, their mean SED was compensated in the FIR with the stacked mean fluxes 
at 250, 350 and 500\,$\mu$m from the SPIRE images.
These values were estimated by first cutting out maps around the MIPS positions for 
individual sources (interpolation into sub-pixels was allowed for precisely centering),
and then measured from the stacked map
via a centered point-spread function (PSF) fitting.
Errors associated with the mean fluxes were calculated using the bootstrap method.
SEDs for both populations resemble the R06/E94 templates in the optical and the NIR,
and differ mainly in the FIR: the stacked $S_{250}$ for FIR-undetected sources is $\sim$ 8\,mJy, 
about 4 times lower than the median $S_{250}$ for FIR-detected quasars (31.1\,mJy).
UV-optical reddening is common in both populations, 
being present in $\sim$ 40\% of the FIR-detected quasars.
The reddening corrections are complicated \citep{hao05} 
and beyond the scope of this paper, as we concentrate in the FIR.

\subsection{Modeling the FIR quasar SED}
\label{sec:fit}
For this work, we adopted a 
T-$\alpha$-$\beta$ model from \citet{blain03} to estimate the dust temperatures and quasar luminosities.
Different from a pure modified blackbody (MBB) model, which uses an exponential thermal function 
with emissivity index $\beta$ to account for a single temperature dust component\footnote{$f_{\rm \nu} \propto {\rm \nu}^{\beta} 
B_{\rm \nu, T}$, where $B_{\rm \nu, T}$ is the blackbody spectrum.};
in the T-$\alpha$-$\beta$ model, a power-law Wien tail ($f_{\rm \nu} \propto {\rm \nu}^{-\alpha} B_{\rm \nu, T}$) is introduced 
to the mid-IR SED to account for the warmer dust components (Fig.~\ref{fig:sed}).
This additional term is then matched to the MBB component at a transition point, 
where the two functions also have equal first order derivatives.
This transition wavelength varies from case to case.
We adopted $\beta = 2.0$ here \citep{priddey03}\footnote{A change 
in $\beta$ from 2.0 to 1.5 does not significantly ($ < 3 \sigma)$ change the 
fitted $T_{\rm d}$ and the $L_{\rm FIR}$ (see also Sec.~\ref{sec:lum}).
Since real dust may have a temperature distribution, 
the fitted temperature only applies to the cold dust component defined by the rest-frame FIR data, 
while the fitted $\alpha$ term indicates the relative strength of warm and hot dust.}.
At shorter wavelengths, 
we normalized the R06 template to each SED over the 1--5\,$\mu$m range
for reference.

Out of the 32 quasars, four only have a single band rest-frame FIR detection. 
One (LHS-M020) of the four sources matches well to the R06 template in the optical and near infrared bands.
The quasar with the highest \textit{z}, LHS--M268, also has an SED that is well described by the R06 template 
alone (within 0.2 dex in the FIR).  
So no fit was attempted for these five objects.

We carried out T-$\alpha$-$\beta$ fits to the remaining 27 SEDs, 
of which 9 objects have all three SPIRE detections. 
Longward of the Lyman break (912\,\AA), 
twenty-two quasars are well-defined by the R06 template
and a T-$\alpha$-$\beta$ fit. 
All 12 SDSS quasars fall into this category.
The remaining five less-well-defined SEDs show a strong stellar bump in the optical-NIR regime, 
presumably from the host galaxy, 
while their FIR SEDs are well described by a T-$\alpha$-$\beta$ fit.
Fig.~\ref{fig:sed} presents the individual SED for each of the 32 FIR-detected quasars. 
The plots are labeled with their ID, redshift, fitted dust temperature and $\alpha$, 
and follow an $\alpha$ decreasing order to match Table.~\ref{tab:param}.
No host galaxy correction was applied.

\subsection{Dust temperatures and luminosities}
\label{sec:lum}
\begin{figure*}
\begin{center}
\includegraphics[scale=0.7]{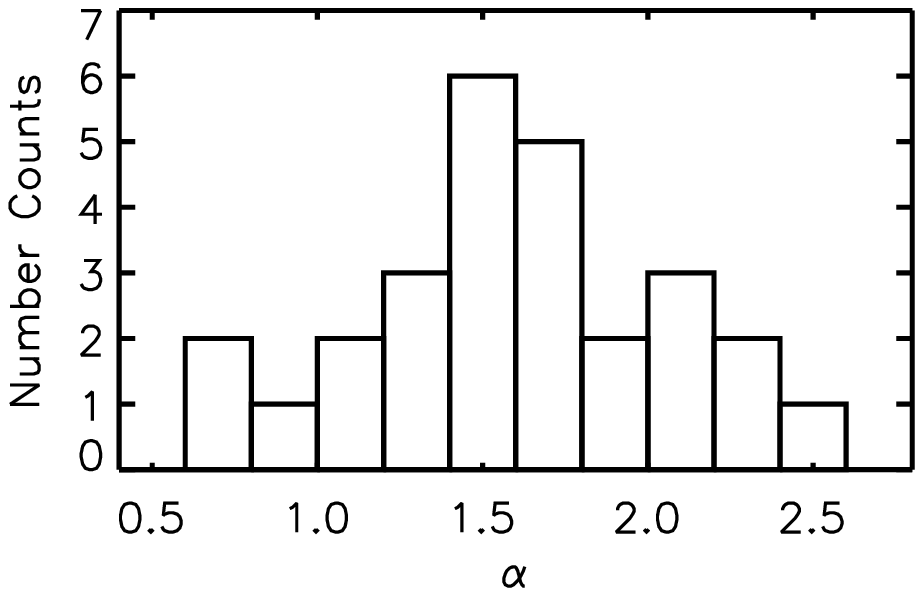}
\includegraphics[scale=0.7]{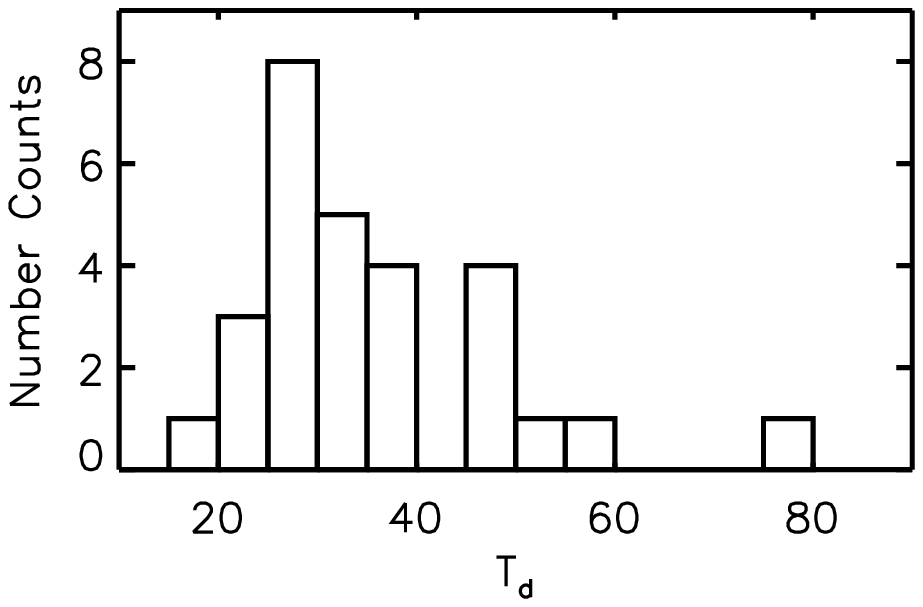}
\end{center}
\figcaption{
Histograms of the fitted power-law slope $\alpha$ (left) and the dust temperature $T_{\rm d}$ (right) for the FIR-detected quasars.
\label{fig:tdust}}
\end{figure*}

The fitted dust temperature ($T_{\rm d}$) for these 27 quasars has a range 
of 18\,K--80\,K, with 87\% of the sources in the range of 25--60\,K, and a median and mean of 
29\,K and 34\,K (Fig.~\ref{fig:tdust}, right).
The fitting error for $T_{\rm d}$ is less than 10\% in $\sim$ 70\% of the cases. 
However, it is worth noting that
the $T_{\rm d}$ derived from T-$\alpha$-$\beta$ fit 
is on average 30$\%$ lower than that from a pure modified blackbody fit, 
where no power-law term is present. 
The FIR luminosities are similar within 3 $\sigma$ between these two fits.

The fitted $\alpha$ for these 27 sources has a wide range of 0.68 to 2.44 (Table.~\ref{tab:param}). 
Starbursts are found to have higher $\alpha$ values than normal star forming galaxies and quasars,
as was shown in \citet{blain03}. 
Different $\alpha$ values suggest different dust temperature compositions in individual 
quasar systems, and may be associated with different evolutionary stages: 
flatter slopes ($\alpha < 1.0$) imply infrared SEDs with relatively stronger warmer dust emissions, 
likely heated directly by the quasar; 
while steeper slopes $\alpha > 2.0$ indicate a colder dust dominant infrared SED, 
similar to that of star forming galaxies (Fig.~\ref{fig:tdust}, left).
Majority of the 27 quasars ($\sim$ 70\%) has an $\alpha$ value of $1.0 \leq \alpha \leq 2.0$,
possibly in a mixed condition between the two extremes.

The FIR luminosities $L_{\rm FIR}$ (40--300\,$\mu$m) were estimated by integrating over the fitted SEDs, 
while the total infrared luminosities $L_{\rm IR}$ (8--1000\,$\mu$m) \citep{kennicutt98} were integrated
over the observed SEDs up to the redshifted MIPS 24\,$\mu$m data, 
and over the fitted SEDs at longer wavelength.
$L_{\rm FIR}$ ranges from $10^{11.3}$ to $10^{13.5}$\,$\lsun$, 
while $L_{\rm IR}$ has a range of $10^{11.5}$ to $10^{14.3}\,$L$_{\sun}$, 
qualifying most of their host galaxies as ultra- or hyper-luminous infrared 
galaxies \citep[ULIRGs, $L_{\rm IR} > 10^{12}\,\lsun$; HyLIRGs, $L_{\rm IR} > 10^{13}\,\lsun$,][]{sm96}:
in the 27 quasars with a T-$\alpha$-$\beta$ fit, there are 8 ULIRGs and 13 HyLIRGs.
Twenty-one of the 27 quasars have $L_{\rm FIR} / L_{\rm IR} > 0.2$, 
and 16 show  $L_{\rm FIR} / L_{\rm IR} > 0.5$, 
indicating major contribution from the FIR to the total $L_{\rm IR}$.
For FIR-undetected quasars,
this ratio is $\sim$0.3 based on their mean SED (Fig.~\ref{fig:sedb}).
We also estimated the `big blue bump' luminosity, $L_{\rm bbb}$ (0.1--0.4\,$\mu$m),
and the near-infrared luminosity, $L_{\rm NIR}$ (2--10\,$\mu$m), by integrating over the observed SEDs
for all quasars. 
The 2--10 $\mu$m range was chosen to minimize the stellar contribution, 
and to better represent the `hot dust bump' emission likely to be directly heated by the quasar \citep{wang08}. 
These integration ranges can be found in Fig.~\ref{fig:sedb} as shaded regions.
The parameters used and derived from the SEDs are summarized in Table.~\ref{tab:param} (See also Fig.~\ref{fig:tdust}).
Errors for $T_{\rm d}$, $\alpha$, $L_{\rm FIR}$ and $L_{\rm IR}$ are fitting errors only.
The errors for $L_{\rm bbb}$ and $L_{\rm NIR}$ contain only the photometric errors,
and are not listed in Table.~\ref{tab:param} because of their small values: 
mostly at the one percent level or less for $L_{\rm bbb}$, 
and at most a few percent for $L_{\rm NIR}$.

The FIR-detected and undetected quasars have similar $L_{\rm bbb}$ and $L_{\rm NIR}$ values and distributions,
with $L_{\rm NIR}$ $\sim$ 25\% higher for FIR-detected quasars at higher redshifts (Fig.~\ref{fig:lratio}, left).
The $L_{\rm FIR} / L_{\rm NIR}$ ratio, however,  shows an obvious excess over the E94 template range
especially at $L_{\rm NIR} \le 10^{13}\,\lsun$ (Fig.~\ref{fig:lratio}, right).
We also plot the mean $L_{\rm FIR} / L_{\rm NIR}$ ratios for FIR-undetected quasars
in three $L_{\rm NIR}$ bins. 
These values were derived from the mean SEDs at each $L_{\rm NIR}$ bin
in the $1< z < 2$ range (covering the mean $z$ of $\sim$1.5), 
assuming $\beta = 2.0$ and $\alpha =1.6$ (mean for FIR-detected quasars).
These SEDs were completed in the FIR using the stacked fluxes from the SPIRE images
for the relevant FIR-undetected sources. 
For comparison, FIR-detected quasars in the same redshift range are marked with a cross in the center. 

It is clear that at the same redshift (1$< z < $ 2 in this case) and luminosity bins, FIR-detected quasars have higher 
$L_{\rm FIR} / L_{\rm NIR}$ ratios than FIR-undetected objects, indicating 
differences in $M_{\rm d}$, $T_{\rm d}$, or AGN activities. 
This ratio seems to decrease at higher $L_{\rm NIR}$ regardless of redshift, 
though the large scatter at $L_{\rm NIR} < 10^{13}\,\lsun$ prevents 
determination of a tight correlation.
The quasars' FIR properties cannot be predicted from shorter wavelengths, 
given the similar SED shapes and luminosities between FIR-detected 
and undetected quasars at optical and near-infrared regions (Fig.~\ref{fig:lratio}, left). 
Considering this similarity and the common UV extinctions, 
no bolometric luminosity was calculated for the FIR-detected quasars. 

\section{DISCUSSION}
\label{sec:dis}
\begin{figure*} 
\begin{center}
\includegraphics[scale=0.7]{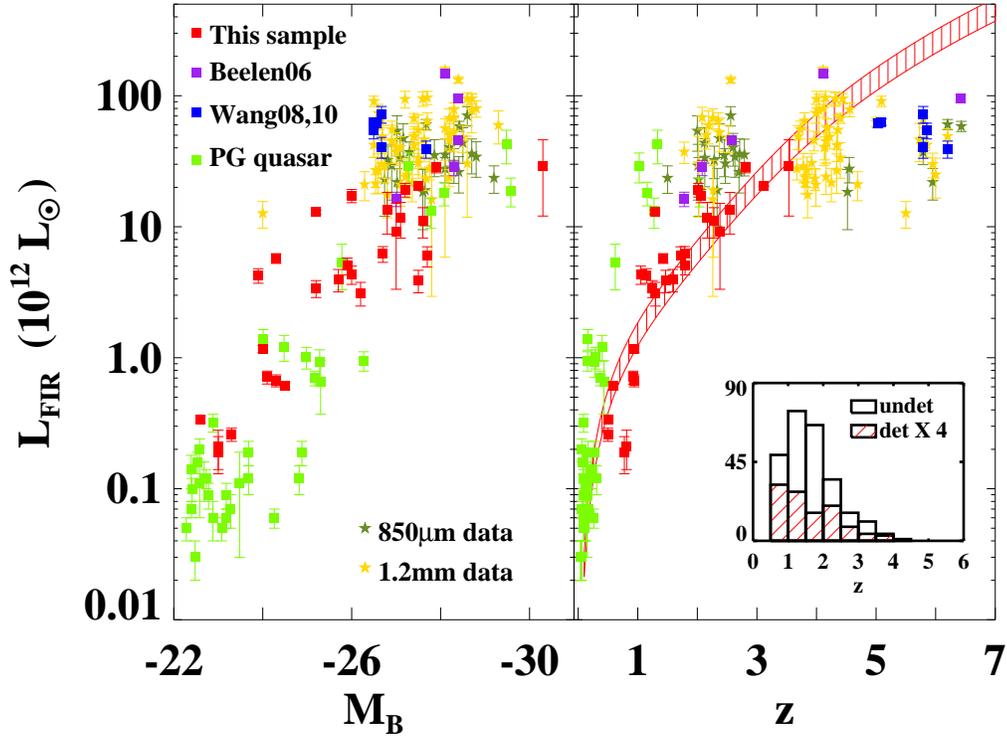}
\end{center}
\figcaption{
Relation between the FIR luminosities and absolute B band magnitudes (left panel), 
and redshifts (right panel) for {\it Herschel} FIR-detected quasars (red),
ISO-detected PG quasars (green),  and (sub)mm-detected quasars --- 
olive and yellow stars for quasars with single band (sub)mm photometry, 
purple and blue squares for those with multi-band (sub)mm data (for references, see Sec.~\ref{sec:SED}). 
 $L_{\rm FIR}$ was estimated with a T-$\alpha$-$\beta$ method for all sources (Sec.~\ref{sec:fit}).
The shaded area in red plots the SPIRE selection function at $S_{250} = 20$\,mJy, 
and the range results from the assumed 20\% flux error (5\,$\sigma$). 
This selection function was derived using fixed $\alpha = 1.6$ and $T_{\rm d} = 34\,K$ 
(mean values of the 27 objects with a T-$\alpha$-$\beta$ fit). 
Some sources are below the SPIRE selection region due to lower $T_{\rm d}$
or higher $\alpha$.
For objects with only one photometric point (850\,$\mu$m or 1.2\,mm data), 
we adopted for the fit $\alpha = 1.6$, and $T_{\rm d} = 43K$ --- mean 
values for (sub)mm quasars with multi-wavelength data \citep{bee06, wang08,wang10}.
The inset shows the redshift distribution for the FIR-undetected quasars in our sample, 
with the 32 FIR-detected quasars in red hatching and scaled by 4 $\times$ for comparison.
\label{fig:submm}}
\end{figure*}

\begin{figure*} 
\begin{center}
\includegraphics[scale=0.8,angle=0]{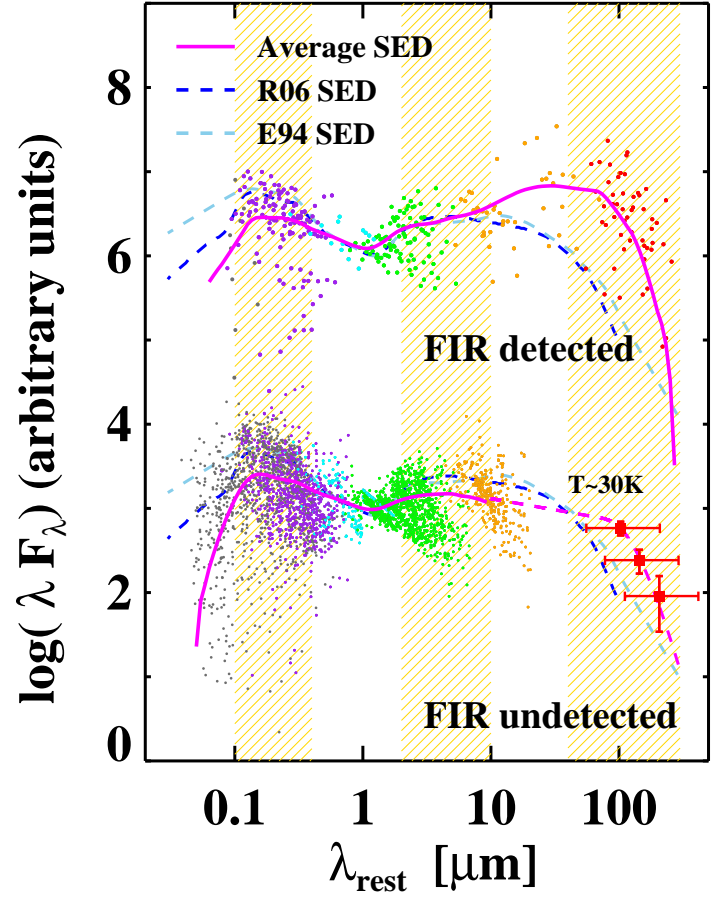}
\end{center}
\figcaption{
Rest-frame mean SEDs for FIR-detected and undetected quasars (arbitrary units).
The colored dots show the detections (normalized to R06 template at 1--5\,$\mu$m): \textit{grey}, GALEX;
\textit{purple}, SDSS; \textit{cyan}, UKIDSS; \textit{green}, SWIRE-IRAC;   
\textit{orange}, SWIRE-MIPS; \textit{red}, HerMES-SPIRE.      
In magenta are the mean SEDs 
for FIR-detected and undetected quasars derived from these normalized detections (see Sec.~\ref{sec:SED}). 
Dashed curves in dark and light blue mark the relative positions of the E94 and R06 templates (normalized at 1\,$\mu$m). 
For FIR-undetected quasars, the stacked mean fluxes from the SPIRE images are plotted as red squares, 
shifted to the rest-frame assuming $z = 1.5$ (mean for FIR-undetected objects),
and errors in the x-axis show the redshift span of 0.2 $\sim$ 4.0.
The regions used for $L_{\rm bbb}$, $L_{\rm NIR}$, and $L_{\rm FIR}$ integration
are shaded in yellow.
\label{fig:sedb}} 
\end{figure*}

\begin{figure*}
\begin{center}
\includegraphics[scale=0.5,angle=0]{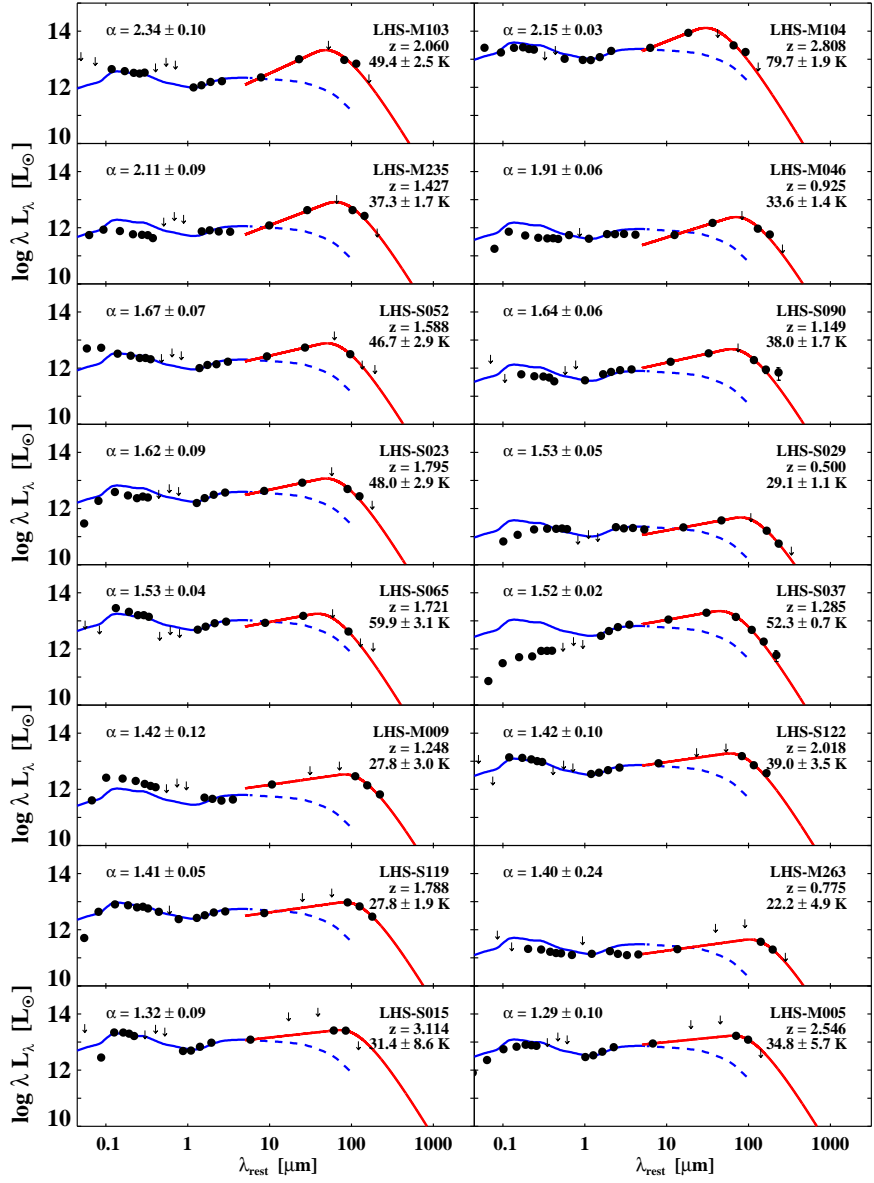}
\end{center}
\figcaption{
Individual rest-frame SEDs for the FIR-detected quasars in our sample, with energy $\lambda L_{\lambda}$ in L${_\sun}$.
The SEDs were constructed with existing data from the UV to the mid-infrared, 
complemented in the FIR by the HerMES-SPIRE observations (see Fig.~\ref{fig:sedb} for details).
The red curves plot a T-$\alpha$-$\beta$ fit with $\beta = 2.0$,
with fitted $T_{\rm d}$ and $\alpha$ given in the legend,
and blue curves are the R06 template normalized at 1--5\,$\mu$m. 
The parameters are summarized in Table.~\ref{tab:param}.
The objects are arranged in an $\alpha$ decreasing order, and followed by quasars with special SED shapes.
No host galaxy correction was applied.
\label{fig:sed}} 
\end{figure*}

We constructed a sample of 32 quasars (0.5 $\leq z <$ 3.6) with {\it Herschel}  SPIRE detections, 
which triples the size of (sub)mm observed quasars 
that have detailed FIR SEDs.
These FIR-detected quasars, as well as some FIR-undetected quasars in our sample (Fig.~\ref{fig:sedb}),
show broad line features and strong cold dust emissions simultaneously.
This is inconsistent with the evolutionary scenario that naked quasars are only seen when the dust has been blown out \citep[e.g.][]{haas03}. 
The dust detected by {\it Herschel} show temperatures from 18\,K to 80\,K (\S\ref{sec:lum}), 
with 87\% of the sources in a range of 25--60\,K,
similar to local and $z\sim2$ starburst galaxies and ULIRGs (20\,K $< T_{\rm d} <$60\,K) \citep{calzetti00, magdis10}.
The median and mean $T_{\rm d}$ in our sample are 29\,K and 34\,K, respectively.  
Since the FIR emissions for these FIR-detected and 
the (sub)mm detected quasars \citep[e.g.][]{omont01, omont03, car01, bee06}
both fall in the rest-frame 30--300\,$\mu$m region,
we associated these two populations as a `dust-rich quasar' population (Fig.~\ref{fig:submm}). 
Estimated dust mass confirmed this connection (See also Sec.~\ref{sec:SED}). 
ISO-detected quasars \citep{haas03} are possibly at the low-{\it z} end of
this population, though their $M_{\rm d}$ could to be 1$\sim$2 dex lower. 
A common assumption is that the dust-rich quasars are in transition 
between optically obscured and unobscured quasar phases, 
but near-infrared/(sub)mm spectroscopy and high resolution images are needed to test this.

\begin{figure*}
\begin{center}
\includegraphics[scale=0.8]{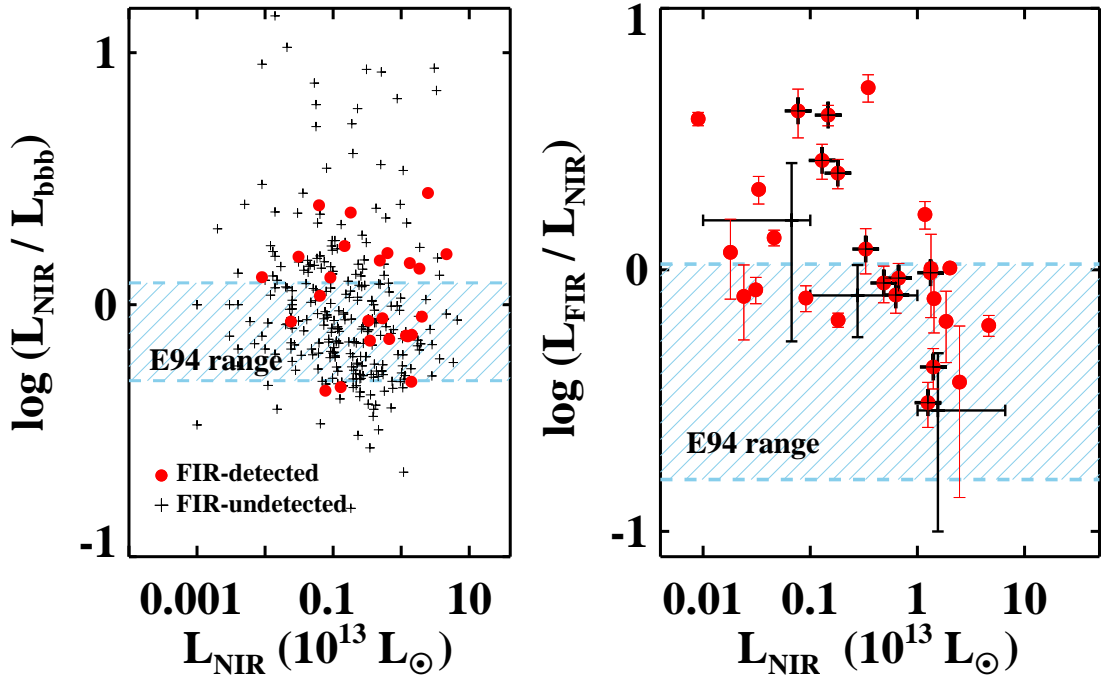}
\end{center}
\figcaption{
{\em(left)} Ratio between the near-infrared and the `big blue bump' luminosities (definitions in
Sec.~\ref{sec:lum} and Fig.~\ref{fig:sed}, left). 
Hatched area marks the E94 predicted range (90\% confidence, E94 was used since R06 template stops at rest-frame 100\,$\mu$m). 
 {\em(right)} The relation between FIR and NIR luminosities for FIR-detected quasars. 
Eleven FIR-detected sources have a higher ratio than the E94 prediction. 
Significant higher $L_{\rm FIR} / L_{\rm NIR}$ ratios are observed at $L_{\rm NIR} \le 10^{13}\,\lsun$, 
whereas more luminous quasars ($L_{\rm NIR} > 10^{13}\,\lsun$) show relatively weaker FIR dust emission. 
Large black crosses are the mean $L_{\rm FIR} / L_{\rm NIR}$ ratios for FIR-undetected quasars at $1< z < 2$
in each $L_{\rm NIR}$ bin.
FIR-detected quasars of the same redshift range are marked with a small cross at the center. 
\label{fig:lratio}}
\end{figure*}

The FIR-detected and undetected quasars have similar mean SEDs (Fig.~\ref{fig:sedb}),
redshift (Fig.~\ref{fig:submm}, inset) and luminosity distributions (Fig.~\ref{fig:lratio}, left) at shorter wavelengths (rest-frame 0.3--20\,$\mu$m).
$L_{\rm NIR}$ is $\sim 25\%$ higher for the FIR-detected population, 
likely affected by their FIR excess. 
The lack of correlation between properties at the FIR and shorter wavelengths
tests the widely used conversion factor of 7 
between $L$(5100\,\AA) and $L_{\rm bol}$ for AGN and quasars \citep{shemmer04}.
For these FIR-detected broad line quasars ($\sim$ 10\% of the flux-limited sample), 
if the rest-frame FIR emission is mainly due to quasar heating,
this factor should be modified to 9$\sim$20 based on the $L_{\rm FIR} / L_{\rm IR}$ ratio. 
Whether this applies to the flux-limited quasar population in general remains to be investigated.

At $1 < z < 2$, the $L_{\rm FIR} / L_{\rm NIR}$ ratios for FIR-detected quasars
are on average $2\times$ higher than FIR-undetected ones. $L_{\rm FIR}/L_{\rm NIR}$ and $L_{\rm NIR}$ 
seem to be anti-correlated (Fig.~\ref{fig:lratio}, right).  
This trend is also observed for the overall population despite the large scatter at $L_{\rm NIR} < 10^{13}\,\lsun$: 
the relative strength of $L_{\rm FIR}$ --- commonly associated with star formation \citep[e.g.][]{lutz07},  
decreases at higher $L_{\rm NIR}$, indicator of warm dust partially or mostly heated by AGN.
This trend is consistent with the assumption that star formation is suppressed 
by the presence of a powerful AGN \citep{hopkins06}.

Both star forming galaxies and AGNs may contribute to the rest-frame FIR emission at various redshifts.
If we attribute the FIR luminosity to star formation\footnote{Eq(4) in \citet{kennicutt98}: 
SFR ($\msun yr^{-1}$) = 1.7 $\times 10^{-10} L_{\rm IR} (\lsun)$. 
Here we used $L_{\rm FIR}^{40-300}$ instead of $L_{\rm IR}^{8-1000} $ for a more 
conservative estimate that reduces further contamination from the AGN.},
as was adopted in some previous studies \citep[e.g.][]{evans06, rie06, netzer07,wang10,wang11},  
about 40$\%$ of the sample will require a star formation rate (SFR) $> 1000\,\msun yr^{-1}$, 
similar to that of submillimeter galaxies \citep[e.g.][]{lutz08}. 
However, for the high luminosity end ($L_{\rm NIR} \ge 3 \times 10^{13}\,\lsun$, Fig.~\ref{fig:lratio}), 
with only one source, 
the statistics are insufficient to prove whether there is a similar 
proportion of starburst dominated quasars, as was found for
(sub)mm observed quasars \citep[i.e.$\sim$20-30\%,][]{wang08}.

On the other hand, for some quasars, 
the SFR derived from $L_{\rm FIR}$ reaches 5000\,$\msun yr^{-1}$, 
which is unlikely and probably implies that part of the FIR emission is powered by AGN. 
In Fig.~\ref{fig:lratio} (right), we found that
for the more luminous quasars  ($L_{\rm NIR} > 5 \times10^{12}\,\lsun$), 
the $L_{\rm FIR} / L_{\rm NIR}$ ratio falls in the normal quasar range,
suggesting pure quasar heating. 
Resolved CO and PAH emission from NIR/mm spectroscopy 
and high resolution imaging showing dust distribution will provide a better
estimate of the relative contributions from AGN and starbursts
for these dust-rich quasars. 

\acknowledgments{
This research has made use of data from the HerMES project --- a {\it Herschel} Key Program 
utilizing Guaranteed Time from the SPIRE instrument team, ESAC scientists and a mission scientist. 
SPIRE has been developed by a consortium of institutes led
by Cardiff Univ. (UK) and including Univ. Lethbridge (Canada);
NAOC (China); CEA, LAM (France); IFSI, Univ. Padua (Italy);
IAC (Spain); Stockholm Observatory (Sweden); Imperial College
London, RAL, UCL-MSSL, UKATC, Univ. Sussex (UK); Caltech,
JPL, NHSC, Univ. Colorado (USA). This development has been
supported by national funding agencies: CSA (Canada); NAOC
(China); CEA, CNES, CNRS (France); ASI (Italy); MCINN (Spain);
SNSB (Sweden); STFC and UKSA (UK); and NASA (USA).
The HerMES data were accessed through the HeDaM database (http://hedam.oamp.fr) 
operated by CeSAM and hosted by the Laboratoire d'Astrophysique de Marseille.
We acknowledge support from the Science and Technology Facilities Council [grant number ST/F002858/1] and [grant number ST/I000976/1].
This work is based partly on observations made with the {\it Spitzer} Space Telescope and the MMT Observatory,
operated by the Jet Propulsion Laboratory, Caltech under a contract with NASA,
and the Smithsonian Astrophysical Observatory and the University of Arizona, respectively.
Research by Y. S. D is supported by the SAO Predoctoral Fellowship.}

{\it Facility:} \facility{Herschel Space Telescope}, \facility{MMT},  \facility{Spitzer Space Telescope}

\begin{figure*}
\begin{center}
\includegraphics[scale=0.6,angle=0]{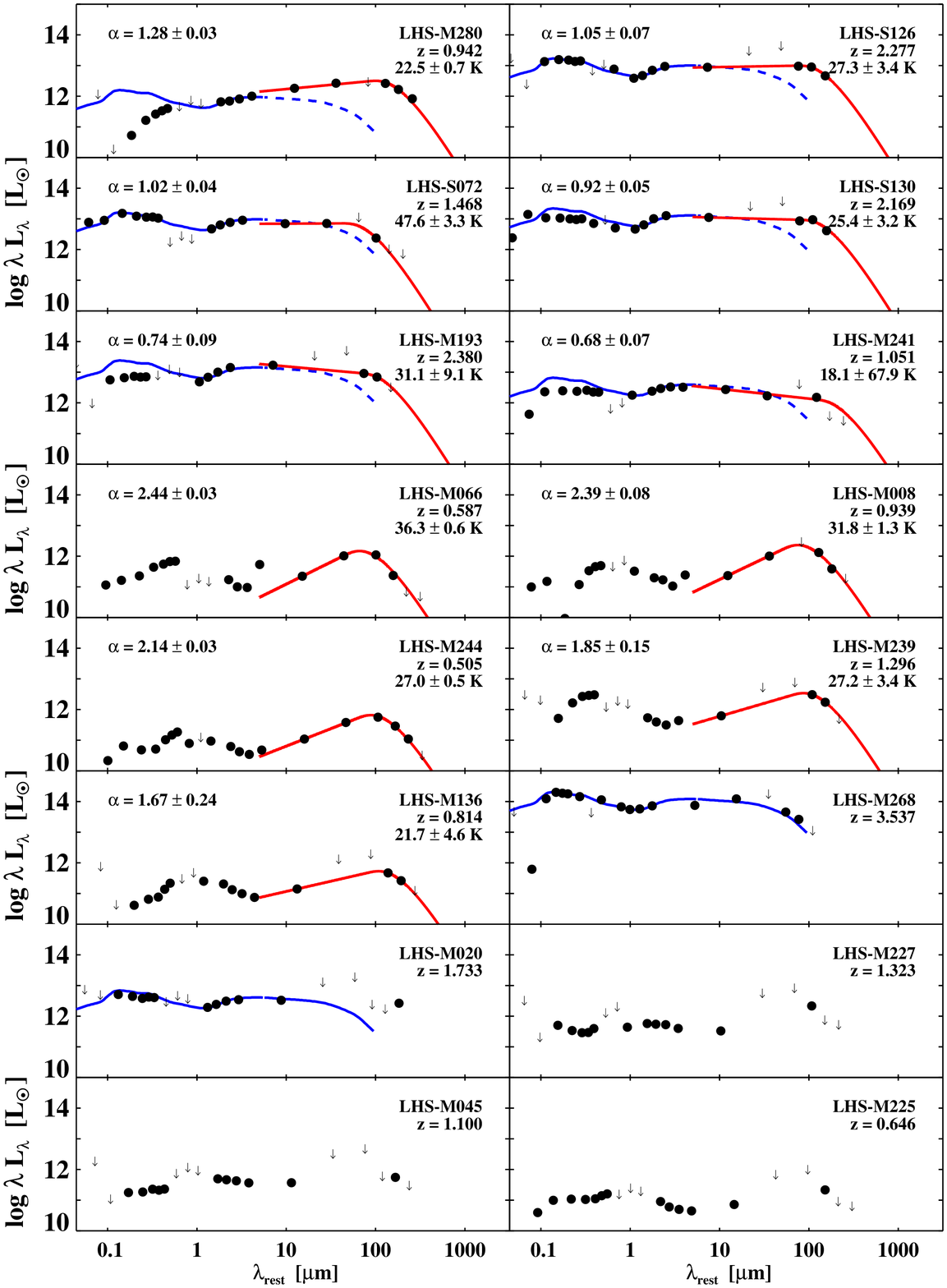}
\end{center}
\caption{Fig.~\ref{fig:sed}, Continued.
\label{fig:fig4b}}
\end{figure*}

\begin{deluxetable*}{lcccccccccc}
\tablecaption{Parameters derived from the fitted SED.  \label{tab:param}}
\tablehead{Object ID & z  & $T_{\rm d}$& $\alpha$ & $L_{\rm bbb}^{0.1-0.4\,\mu m}$ & $L_{\rm NIR}^{2-10\,\mu m}$ & $L_{\rm FIR}^{40-300\,\mu m}$ &$L_{\rm IR}^{8-1000\,\mu m}$ &  SFR  & log$M_{\rm d}$ \\
  && (K)&& (10$^{12} \,\lsun$)&(10$^{12} \,\lsun$)&(10$^{12} \,\lsun$)&(10$^{12} \,\lsun$)& ($M_{\sun} / $yr) & (M$_{\sun}$) \\
(1) & (2) &(3) & (4) &(5) & (6) & (7) & (8) & (9) &(10)} 
\startdata 
LHS-M103  &  2.060   &     49.4$\pm$2.5   &  2.34$\pm$0.10    &     4.82    &       3.47      &     17.26$\pm$2.01     &      31.00$\pm$3.61     &     3000        &    8.9     \\
LHS-M104  &  2.808   &     79.7$\pm$1.9   &  2.15$\pm$0.03    &    29.18    &      46.30      &     28.38$\pm$2.59     &     115.84$\pm$10.56    &     5000        &    8.9     \\
LHS-M235  &  1.427   &     37.3$\pm$1.7   &  2.11$\pm$0.09    &     0.86    &       1.47      &      5.74$\pm$0.49     &       8.68$\pm$0.74     &     1000        &    9.1     \\
LHS-M046  &  0.925   &     33.6$\pm$1.4   &  1.91$\pm$0.06    &     0.71    &       0.91      &      0.71$\pm$0.08     &       1.05$\pm$0.12     &     100         &    9.4     \\
LHS-S052  &  1.588   &     46.7$\pm$2.9   &  1.67$\pm$0.07    &     3.81    &       3.29      &      3.95$\pm$0.77     &       8.00$\pm$1.57     &     700         &    8.6     \\
LHS-S090  &  1.149   &     38.0$\pm$1.7   &  1.64$\pm$0.06    &     0.78    &       1.81      &      4.24$\pm$0.54     &       7.29$\pm$0.93     &     700         &    8.7     \\
LHS-S023  &  1.795   &     48.0$\pm$2.9   &  1.62$\pm$0.09    &     3.92    &       6.28      &      5.03$\pm$0.74     &      10.82$\pm$1.60     &     900         &    9.5     \\
LHS-S029  &  0.500   &     29.1$\pm$1.1   &  1.53$\pm$0.05    &     0.20    &       0.31      &      0.26$\pm$0.03     &       0.39$\pm$0.04     &     40          &    8.7     \\
LHS-S065  &  1.721   &     59.9$\pm$3.1   &  1.53$\pm$0.04    &    28.45    &      14.07      &      5.98$\pm$1.05     &      16.93$\pm$2.97     &     1000        &    8.5     \\
LHS-S037  &  1.285   &     52.3$\pm$0.7   &  1.52$\pm$0.02    &     0.81    &      13.26      &     12.92$\pm$0.50     &      29.86$\pm$1.15     &     2200        &    8.2     \\
LHS-M009  &  1.248   &     27.8$\pm$3.0   &  1.42$\pm$0.12    &     2.74    &       1.29      &      3.38$\pm$0.50     &       5.69$\pm$0.84     &     600         &    9.1     \\
LHS-S122  &  2.018   &     39.0$\pm$3.5   &  1.42$\pm$0.10    &    15.60    &      11.76      &     19.14$\pm$2.30     &      39.32$\pm$4.73     &     3300        &    9.3     \\
LHS-S119  &  1.788   &     27.8$\pm$1.9   &  1.41$\pm$0.05    &     9.13    &       6.67      &      6.20$\pm$0.85     &      10.94$\pm$1.49     &     1100        &    9.4     \\
LHS-M263  &  0.775   &     22.2$\pm$4.9   &  1.40$\pm$0.24    &     0.28    &       0.24      &      0.19$\pm$0.06     &       0.30$\pm$0.09     &     30          &    8.6     \\
LHS-S015  &  3.114   &     31.4$\pm$8.6   &  1.32$\pm$0.09    &     22.42    &      20.10      &     30.16$\pm$21.82    &     58.81$\pm$42.55     &     3500        &    8.7     \\
LHS-M005  &  2.546   &     34.8$\pm$5.7   &  1.29$\pm$0.10    &     9.15    &      13.38      &     13.54$\pm$4.75     &      27.99$\pm$9.82     &     2300        &    8.8     \\
LHS-M280  &  0.942   &     22.5$\pm$0.7   &  1.28$\pm$0.03    &     0.09    &       1.82      &      1.17$\pm$0.07     &       1.97$\pm$0.12     &     200         &    8.5     \\
LHS-S126  &  2.277   &     27.3$\pm$3.4   &  1.05$\pm$0.07    &    18.82    &      14.29      &     11.09$\pm$2.89     &      24.44$\pm$6.37     &     1900        &    9.3     \\
LHS-S072  &  1.468   &     47.6$\pm$3.3   &  1.02$\pm$0.04    &    16.73    &      12.53      &      3.89$\pm$0.76     &      10.66$\pm$2.08     &     700         &    8.5     \\
LHS-S130  &  2.169   &     25.4$\pm$3.2   &  0.92$\pm$0.05    &    13.33    &      18.53      &     11.77$\pm$3.58     &      28.01$\pm$8.51     &     2000        &    9.4     \\
LHS-M193  &  2.380   &     31.1$\pm$9.1   &  0.74$\pm$0.09    &     8.91    &      24.72      &      9.20$\pm$5.87     &      26.93$\pm$17.20    &     1600        &    9.3     \\
LHS-M241  &  1.051   &     18.1$\pm$67.9  &  0.68$\pm$0.07    &     3.24    &       4.85      &      2.21$\pm$22.06    &       5.21$\pm$51.91    &     400         &    ...    \\
LHS-M066  &  0.587   &     36.3$\pm$0.6   &   2.44$\pm$0.03   &     0.36    &     0.46$^*$    &      0.61$\pm$0.04     &       0.81$\pm$0.05     &     100         &    8.1     \\
LHS-M008  &  0.939   &     31.8$\pm$1.3   &   2.39$\pm$0.08   &     0.19    &     0.33$^*$    &      0.67$\pm$0.07     &       0.89$\pm$0.09     &     100         &    8.8     \\
LHS-M244  &  0.505   &     27.0$\pm$0.5   &   2.14$\pm$0.03   &     0.07    &     0.09$^*$    &      0.34$\pm$0.02     &       0.44$\pm$0.03     &     60          &    9.1     \\
LHS-M239  &  1.296   &     27.2$\pm$3.4   &   1.85$\pm$0.15   &     1.69    &     0.77$^*$    &      3.12$\pm$0.64     &       4.52$\pm$0.92     &     500         &    8.9     \\
LHS-M136  &  0.814   &     21.7$\pm$4.6   &   1.67$\pm$0.24   &     0.05    &     0.18$^*$    &      0.21$\pm$0.07     &       0.31$\pm$0.10     &     40          &    8.8     \\
LHS-M268$^{\dagger}$  &  3.537   &     ...            &      ...          &   208.91    &     129.92      &     ...                &       ...               &     ...         &    ...     \\
LHS-M020  &  1.733   &     ...            &     ...            &    5.99     &    5.29        &     ...               &        ...               &   ...            & ...   \\
LHS-M227  &  1.323   &     ...            &     ...            &    0.59     &    0.64$^*$    &     ...               &        ...               &   ...            &  ...  \\
LHS-M045  &  1.100   &     ...            &     ...            &    0.25     &    0.62$^*$    &     ...               &        ...               &   ...            &  ...  \\
LHS-M225  &  0.646   &     ...            &     ...            &    0.13     &    0.09$^*$    &     ...               &        ...               &   ...            &  ...  \\

\enddata
\tablecomments{(1): Object ID. (2): Spectroscopic redshifts determined from Hectospec and SDSS spectra.  
(3) (4): Dust temperature and power-law index derived from the T-$\alpha$-$\beta$ fit (see also Sec.~\ref{sec:fit}). 
(5) (6) (7) (8): Luminosities estimated from the SED (see also Sec.~\ref{sec:lum}).
(9): Star formation rate estimated using $L_{\rm FIR}^{40-300\,\mu m}$ following the Kennicutt law (1998), see Sec.~\ref{sec:dis}, footnote.
(10): Dust mass calculated using S$_{\rm 250}$, see Sec.~\ref{sec:SED}, footnote.\\
$*$: not corrected for host galaxy contamination.   $\dagger$: SED well defined by R06 template, no T-$\alpha$-$\beta$ fit needed.
}
\end{deluxetable*}


 \end{CJK}
\end{document}